\documentclass[prb,twocolumn,amsmath,amssymb]{revtex4}

\usepackage{graphicx,color}
\usepackage{multirow}

\begin{document}

\title{Permalloy-based carbon nanotube spin-valve}

\author{H. Aurich}
\author{A. Baumgartner}
\email{andreas.baumgartner@unibas.ch}
\author{F. Freitag}
\author{A. Eichler}
\author{J. Trbovic}
\author{C. Sch\"onenberger}
\affiliation{
Institute of Physics, University of Basel, Klingelbergstrasse 82, 4056 Basel, Switzerland\\}

\begin{abstract}
In this Letter we demonstrate that Permalloy (Py), a widely used Ni/Fe alloy, forms contacts to carbon nanotubes (CNTs) that meet the requirements for the injection and detection of spin-polarized currents in carbon-based spintronic devices. We establish the material quality and magnetization properties of Py strips in the shape of suitable electrical contacts and find a sharp magnetization switching tunable by geometry in the anisotropic magnetoresistance (AMR) of a single strip at cryogenic temperatures. In addition, we show that Py contacts couple strongly to CNTs, comparable to Pd contacts, thereby forming CNT quantum dots at low temperatures. These results form the basis for a Py-based CNT spin-valve exhibiting very sharp resistance switchings in the tunneling magnetoresistance, which directly correspond to the magnetization reversals in the individual contacts observed in AMR experiments.
\end{abstract}

\maketitle

The generation and detection of spin-polarized currents are fundamental building blocks in spintronic devices.\cite{Wolf_Awschalom_Roukes_Science294_2001, Moodera_PhysicsToday63_2010} The most basic structure is known as spin-valve, where the device resistance depends on the relative orientation of the magnetizations in two ferromagnetic (F) contacts coupled to a structure in-between. For a tunnel barrier this effect is known as tunneling magnetoresistance (TMR). Due to the expected large spin relaxation times, carbon nanotubes (CNTs) and graphene are ideal coupling media for such spintronic devices. The F contacts can also be connected to a CNT quantum dot (QD), which allows one to electrically tune the spin-valve effect\cite{Sahoo_Kontos_Schoenenberger_naturephys_2005} and to investigate spin-related phenomena in nanostructures, which is particularly interesting in materials with exotic electronic properties like CNTs or graphene.

In recent research the ferromagnetic materials used to contact CNTs were mainly NiPd,\cite{Sahoo_Kontos_Schoenenberger_naturephys_2005, Feuillet-Palma_Kontos_PRB81_2010} Co\cite{Tsukagoshi_Nature401_1999, Tombros_van_der_Molen_van_Wees_PRB73_2006} and various others.\cite{Jensen_Nygard_Lindelof_PRB72_2005, Thamankar_APL89_2006, Gunnarsson_Trbovic_Schoenenberger_PRB77_2008} In modern technology, however, the materials of choice for applications are Ni/Fe alloys, e.g. Ni$_{80}$Fe$_{20}$ called Permalloy (Py).\cite{Parkin_Science320_2008} Here we show that Py is well-suited as ferromagnetic material in nanoscale spintronic devices and we demonstrate clear TMR switching in a Py-based CNT spin-valve, directly related to the magnetization reversal in the individual F contacts observed in anisotropic magnetoresistance (AMR) measurements.

Ideally, the following requirements have to be met by the ferromagnetic material in a spin-valve: 1) a high spin polarization to obtain a large spin signal, 2) the shape anisotropy dominates the crystal anisotropy which allows one to tune the switching fields by the contact geometry, 3) a single domain covering each contact area to avoid compensation by neighboring domains, 4) in-plane magnetization to reduce stray field effects, and 5) reproducible electronic coupling to the medium between the contacts. For many materials the local magnetization and the contact resistance to carbon nanostructures were found to be of poor reproducibility or to develop an out-of-plane magnetization and multi-domain structures.\cite{Preusche_Strunk_JAP106_2009}

Py can exhibit a large spin-polarization similar to Co or Fe.\cite{Moodera_AnnuRevMaterSci_1999} In addition, it has a small crystal anisotropy and forms only few large domains in nanostructured thin films.\cite{Preusche_Strunk_JAP106_2009} In our fabrication process we first establish the material quality by investigating a thermally evaporated film of the required thickness in a vibrating sample magnetometer (VSM) at room temperature. In this step we find that Py films in the thickness range of $20\,$nm to $50\,$nm are simple to fabricate and are described well by bulk parameters. This test can be done with only minimal preparation.

Because the magnetization of a single contact is too small to be detected by a VSM, we investigate grids of up to $\sim400,000$ Py strips, each with the geometry of a single contact. An SEM image of part of such a sample is shown at the lower right of Fig.~1(a). All Py contacts discussed in this paper are $25\,$nm thick and between $10\,\mu$m and $17\,\mu$m long rectangles. Three magnetization curves with the external magnetic field $H$ applied along the strips are plotted in Fig.~1(a) for strips of $110\,$nm, $220\,$nm and $550\,$nm width. The metal-covered area is approximately the same for each sample, leading to similar saturation magnetizations.

We find that narrow contacts result in a large coercive field, dictated by the shape anisotropy. Except for the $110\,$nm strips, the magnetization reversal from $10\%$ to $90\%$ of the saturation values takes place on $\sim10\,$mT. We denote the field where the magnetization reaches $90\%$ as $H_{90}$. We attribute the different curve shape and the wider transition of the most narrow strips to stray-field interactions between the strips and to an enhanced sensitivity of the magnetization to variations in the lithographic widths (see below).

\begin{figure}[t]{
\centering
\includegraphics{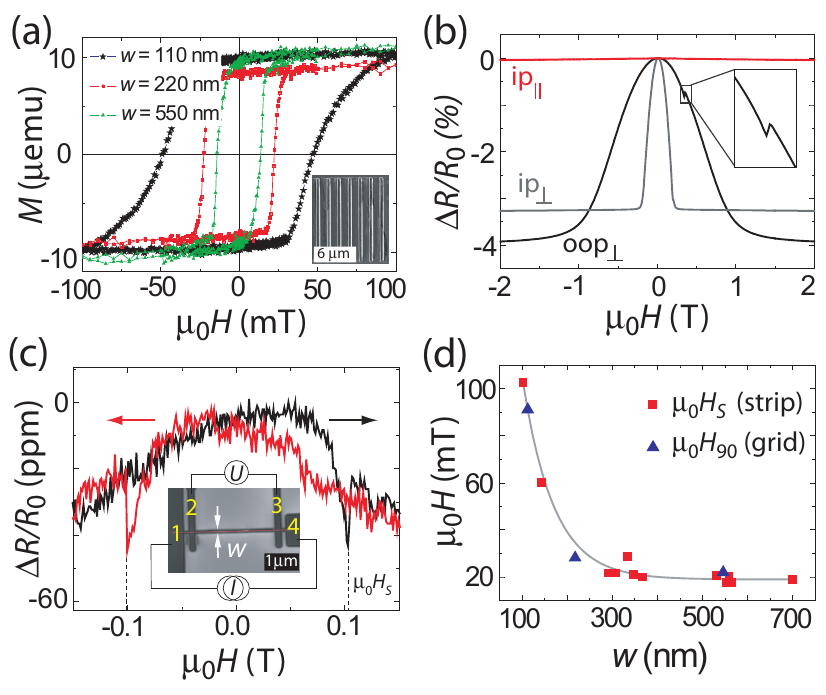}
}
\caption{(Color online) (a) VSM magnetization curves of up to $\sim400,000$ Py strips of the indicated widths. Inset: SEM image of part of such a grid. (b) Four-terminal AMR curves of a single $100\,$nm wide strip (up-sweep only). ip$_{\|}$, ip$_{\bot}$ and oop$_{\bot}$ stand for in-plane parallel, in-plane perpendicular and out-of plane perpendicular, i.e. the magnetic field direction relative to the sample plane and the strip axis. (c) Small-field ip$_{\|}$ AMR signal. The arrows indicate the sweep directions. Inset: SEM image of the sample and measurement schematic. (d) AMR switching field $H_{s}$ and VSM transition field $H_{90}$ as a function of the contact width for $H$ parallel to the strips. The full line is a guide to the eye.
\label{Figure1}}
\end{figure}

Due to spin-orbit interactions the electrical resistance in a ferromagnetic metal depends on the orientation of the magnetization to the current direction. We use this so-called anisotropic magnetoresistance (AMR) to investigate an individual contact strip in a four-terminal geometry with normal metal contacts. The inset of Fig.~1(c) shows an SEM image and the measurement schematic of such a device. In Fig.~1(b) we present the AMR of a single $100\,$nm wide strip as the normalized deviation from the maximum value $R_{0}$ at a temperature of $4.2\,$K. The magnetic field is aligned parallel (ip$_{\|}$) and perpendicular to the strip, the latter either in the sample plane (ip$_{\bot}$), or orthogonal to the surface (oop$_{\bot}$). While the AMR hardly changes for ip$_{\|}$, the curves recorded with the field perpendicular to the strip both show the characteristic shape of a continuous rotation of a magnetic moment with increasing field. The maximum signal change of $3\%-4\%$ is consistent with bulk AMR measurements.\cite{McGuire_TransctionsMagnetics11_1975} The data for the reversed sweep direction are omitted for clarity and coincide with the presented curves. The switching due to a small misalignment shown in the inset is considerably smaller than the signal change due to the magnetization rotation and occurs at the corresponding negative field in the down-sweep. The differences between ip$_{\bot}$ and oop$_{\bot}$ are probabely due to the different demagnetization factors and domain structures in the two directions. These AMR curves are well-described by the Stoner-Wohlfarth model for the coherent rotation of a single macroscopic magnetic moment with the anisotropy energy as a fit parameter and the relation $\Delta R(H)=\Delta R_{\rm max}\cos(\Theta(H))$, with $\Theta$ the angle between the current and the magnetization. These results suggest that the Py strip contains a large domain with the magnetization directed along the strip axis which dominates the AMR signal.

For a spin-valve the ideal magnetization dynamics is a sharp switching by $180^{\circ}$ at the switching field $H_{\rm s}$ without prior rotation and formation of multiple domains when the magnetic field is reversed. In this case no variation of the AMR is expected. This can be achieved in the ip$_{\|}$ configuration, where a slight misalignment ($\leq 2^{\circ}$) allows one to detect the switching, as shown in Fig.~1(c). One finds a clear discontinuous change in the resistance and a small background, which suggests only a small initial rotation of the magnetization and a very sharp switching on a field scale of $\mu_{0}\Delta H<1\,$mT.

The switching field $H_{\rm s}$ as a function of the width $w$ obtained in AMR measurements of Py strips parallel to the magnetic field is plotted in Fig.~1(d). In addition, also the corresponding VSM transition fields $H_{90}$ are given in the same graph. In the range between $w=100\,$nm and $w=300\,$nm the switching field varies by a factor $5$, or by at least $80$ times the width of a single AMR switching. We conclude that by controlling the width of a contact one can reproducibly tune the switching field on technically and experimentally useful scales.

To demonstrate that Py forms good electrical contacts to CNTs we have fabricated devices as shown in Fig.~2(a): a CNT is contacted in the center by two Py contacts ($2$ and $3$) of width $140\,$nm and $370\,$nm, with an edge to edge distance of $300\,$nm. At the top and bottom two Pd contacts ($1$ and $4$) are placed for comparison since Pd forms good contacts to CNTs.\cite{Mann_Nanolett3_2003} Each Py strip is contacted in a four-terminal geometry for AMR characterization. All structures are fabricated by standard electron beam lithography on highly doped Si substrates serving as backgate with $400\,$nm thermally oxidized SiO$_2$ on top.

At room temperature we find two-terminal resistances of two Py contacts connected by a CNT in the range of $20\,$k$\Omega$ to $50\,$k$\Omega$, similar to our Pd contacts. At $T=230\,$mK the four contacts divide the CNT into three quantum dots labeled QD1, QD2 and QD3 in Fig.~2(a). The upper part of Fig.~2(b) shows clear Coulomb blockade oscillations as a function of $V_{\rm BG}$ in the two-terminal differential conductance of QD1 between contact $1$ (Pd) and $2$ (Py), and for QD2 between the two Py contacts $2$ and $3$, providing evidence for carrier confinement in these regions. The peak values in $G_{\rm QD2}$ reach $0.6\,G_0$ ($G_0=2e^2/h$), suggesting a relatively symmetric coupling. We characterize the coupling of a single Py contact by applying a voltage to contact $1$ and measuring simultaneously the two resulting currents in the Py terminals $2$ and $3$, see Fig.~2(a). The differential conductance between contacts $1$ and $3$, $G_{31}=I_3/U$, is shown in the lower part of Fig.~2(b) and exhibits peak maxima of $\sim0.004\,G_0$, about $30$ times smaller than $G_{21}$, indicating that most of the current flows into contact $2$. As shown in the lower part of Fig.~2(b), the transmission $T_{31}=G_{31}/G_0$ is reproduced very well by the product of the transmissions obtained from the two-terminal conductances of QD1 and QD2, $T_{31}\approx fT_{\rm QD1}T_{\rm QD2}$, with $f\approx0.008$. Such characteristics might be found for strongly decoupled coherent subsystems, i.e. the QDs, where $f$ describes the small transmission between the dots.

\begin{figure}[t]{
\centering
\includegraphics{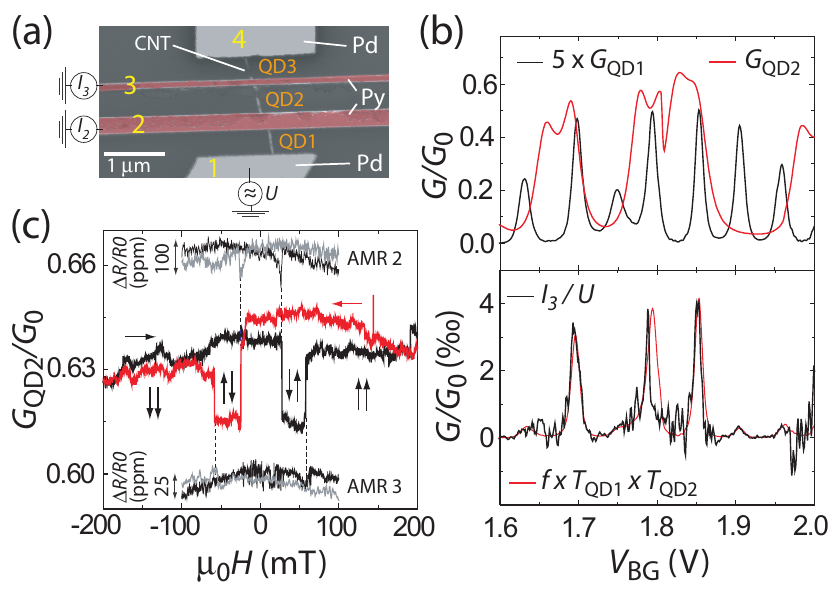}
}
\caption{(Color online) (a) SEM image of a CNT-based spin-valve device with two additional Pd contacts. (b) Coulomb blockade oscillations in the differential conductance as a function of the backgate voltage $V_{\rm BG}$ for QD2 (with two Py contacts) and QD1 at $T=230\,$mK. The bottom part shows the conductance between contacts $1$ and $3$ when contact $2$ is on ground. (c) TMR signal between the two Py contacts on a Coulomb oscillation maximum as a function of the external magnetic field applied parallel to the contacts. The top and bottom curves are the AMR signals of the individual Py contacts 2 (AMR 2) and 3 (AMR 3).
\label{Figure2}}
\end{figure}

Applying the bias on contact 4 (not shown) through the more transparent QD3 and measuring the conductances $G_{43}$ and $G_{42}$ yields only a $\sim8$ times smaller current in contact $2$ than in contact $3$. This finding is consistent with a weaker coupling of the CNT to the narrower Py contact.

The two Py contacts and QD2 form a lateral CNT spin-valve structure. An example of a TMR measurement is plotted in Fig.~2(c). The differential conductance of QD2 between the Py contacts is shown as a function of the external magnetic field applied along the Py strips and at the backgate voltage $V_{\rm BG}=0.656\,$V, corresponding to a Coulomb oscillation maximum. We find clear TMR switchings at $\mu_0 H=\pm26\,$mT and $\mu_0 H=\pm58\,$mT. The relative magnetization orientations are indicated in the figure. These switching fields correspond very well to the fields of the magnetization reversal in the Py contacts found in the AMR measurements plotted above and below the TMR curves. The magnitude of the TMR switching is about $\sim3\%$ and depends strongly on $V_{\rm BG}$.\cite{Sahoo_Kontos_Schoenenberger_naturephys_2005} Using Juli\`ere's model\cite{Julliere_PLA54_1975} and the assumption of equally polarized contacts one finds a contact polarization of $\sim13\%$, which is smaller than expected possibly due to the strong coupling of Py to the CNT.

In summary, we have demonstrated that Py contacts to carbon nanotubes can be engineered by standard methods to meet the requirements of carbon-based spintronic devices. In particular, we have shown very sharp TMR switching on a CNT spin-valve, clearly related to the magnetization switching observed in AMR experiments. This method of comparing AMR and TMR experiments offers additional means to exclude other mechanisms that can lead to signals similar to TMR, e.g. hopping transport between granular islands in the contacts, stray field effects, or the magneto-Coulomb effect. In addition, we have demonstrated that the Py coupling to the CNT is strong and similar to that of Pd contacts.

This work is supported by the Swiss National Science Foundation in the framework of the NCCR Nano.

\bibliographystyle{apsrev}

\end{document}